\documentclass[sigconf]{acmart}

\usepackage{booktabs} 

 \renewcommand\footnotetextcopyrightpermission[1]{} 
\usepackage{mathtools}

\usepackage{arydshln,leftidx}
\usepackage{amsmath}
\usepackage{amssymb}
\usepackage{kbordermatrix}
\usepackage{graphics}
\usepackage{graphicx}
\usepackage{enumitem}
\usepackage{textcomp}
\usepackage{color}
\usepackage{outlines}

\usepackage{listings}
\definecolor{dkgreen}{rgb}{0,0.6,0}
\definecolor{gray}{rgb}{0.5,0.5,0.5}
\definecolor{mauve}{rgb}{0.58,0,0.82}
\usepackage{amsthm}

\usepackage{caption}
\usepackage{setspace}
\usepackage{threeparttable}

\usepackage{changepage}

\usepackage{etoolbox}
\let\bbordermatrix\bordermatrix
\patchcmd{\bbordermatrix}{8.75}{4.75}{}{}
\patchcmd{\bbordermatrix}{\left(}{\left|}{}{}
\patchcmd{\bbordermatrix}{\right)}{\right|}{}{}

\usepackage{blkarray}
\usepackage{multirow}
\usepackage{wrapfig}
\usepackage{tikz}

\usepackage{subfig}
\usetikzlibrary{matrix,fit}

\begin{document}
\title{A Lightweight McEliece Cryptosystem Co-processor Design \vspace{-0.1in}}

\author{Lake Bu, Rashmi Agrawal, Hai Cheng, and Michel A. Kinsy\\
Adaptive and Secure Computing Systems Laboratory\\
Department of Electrical and Computer Engineering, Boston University\\
(bulake, rashmi23, chenghai, mkinsy)@bu.edu}

\begin{abstract}
Due to the rapid advances in the development of quantum computers and their susceptibility to errors, 
there is a renewed interest in error correction algorithms. In particular, error correcting code-based 
cryptosystems have reemerged as a highly desirable coding technique. This is due to the fact that most classical asymmetric cryptosystems will fail in the quantum computing era. Quantum computers can solve many of the integer factorization and discrete logarithm problems efficiently. However, code-based cryptosystems are still secure against quantum computers, since the decoding of linear codes remains as NP-hard even on these computing systems. 
One such cryptosystem is the McEliece code-based cryptosystem. The original McEliece code-based cryptosystem uses binary Goppa code, which is known for its good code rate and error correction capability. However, its key generation and decoding procedures have a high computation complexity. In this work we propose a design and hardware implementation of an public-key encryption and decryption co-processor based on a new variant of McEliece system. This co-processor takes the advantage of the non-binary Orthogonal Latin Square Codes to achieve much smaller computation complexity, hardware cost, and the key size. 
\end{abstract}

\vspace{-0.05in}

\maketitle

\vspace{-0.05in}
\section{Introduction}
\vspace{-0.05in}
Since the possibility of using quantum effects in computation was brought up by Feynman in 1959, numerous efforts have been dedicated to realize and even commercialize the quantum computers. In the past three years, a number of significant milestones have been reached in this area. 
From late 2017 to early 2018, technology companies such as IBM, Intel, and Google, announced their construction and testing of 50-, 49-, and 72-qubit computers, respectively. In July 2018, for the first time, researchers at the University of Sydney successfully realized a multi-qubit computation on a system of trapped ions, which is believed to be the leading platform in building general quantum computers \cite{US2018}. 
In December 2018, IonQ claimed to have built a quantum computer with 160 qubits. Besides these advances in physical implementation of quantum computers, key breakthroughs in the verification of quantum computation were also achieved \cite{UM2017}, and more efficient error correction schemes were recently proposed as well.
These efforts are rooted in the fact that quantum computers promise greater computational power. But these developments also bring with them burning security concerns. For example, Shor's algorithm \cite{PS1999}, leveraging quantum Fourier transform, is able to solve the integer factorization problem efficiently. Therefore, 
current popular cryptographic algorithms such as RSA, ElGamal, Diffie-Hellman, and ECC, which rely on the hardness of integer factorization and discrete logarithm, are vulnerable to quantum computer-based algorithms. 

In response to the aforementioned security challenges associated with quantum computers, a number of new cryptosystems has been proposed for the post-quantum era. In early 2017, NIST launched a campaign for a post-quantum cryptography standardization. Up to February 2019, totally 27 out of 69 candidates of this new standard made to the second round of competition \cite{NIST2019}. Among the 27, the two most likely contenders are the lattice-based cryptosystems (12 candidates), and the code-based cryptosystems (8 candidates). Both of them are able to construct public-key cryptosystems and key exchange mechanisms.

Compared to the popular lattice-based ring-learning with error (Ring-LWE) cryptosystem, error-correcting code (ECC)-based schemes have a much larger key size, which is considered a drawback. However, they do bear the advantage of withstanding the test of time. For example, since its formulation by Robert McEliece in 1978 \cite{ME1978}, the McEliece code-based technique has so far proven to be cryptanalysis resistant (although sometimes increasing the key size is necessary). 
The conventional McEliece cryptosystem uses the binary Goppa code, which has good code rate and error correction capability. However, comparing with other binary codes, the generating and decoding (error-correction) of binary Goppa codes have relatively high complexity since they involve intensive computations over finite fields, including modulo polynomial operations. In addition, the key size of McEliece systems is usually large. For a binary Goppa code with an $k \times n$ generating matrix, the key size is $kn$ with $k, n$ in thousands of bits. 

Therefore, to address these issues, we propose a new variant of McEliece cryptosystem and its encryption-decryption co-processor. The proposed system is based on the generalized non-binary Orthogonal Latin Square Code (OLSC), which is known for its simple encoding and decoding algorithms, leading to potential efficient hardware implementation. In addition, the non-binary OLSC is able to work with non-binary messages through binary matrices. In other words, a long message is able to be processed by relatively small matrices, which reduces the key size. 

\section{The OLSC-based Cryptosystem}

\begin{figure*}[!h]
	\begin{center}
		\includegraphics[width=6.8in]{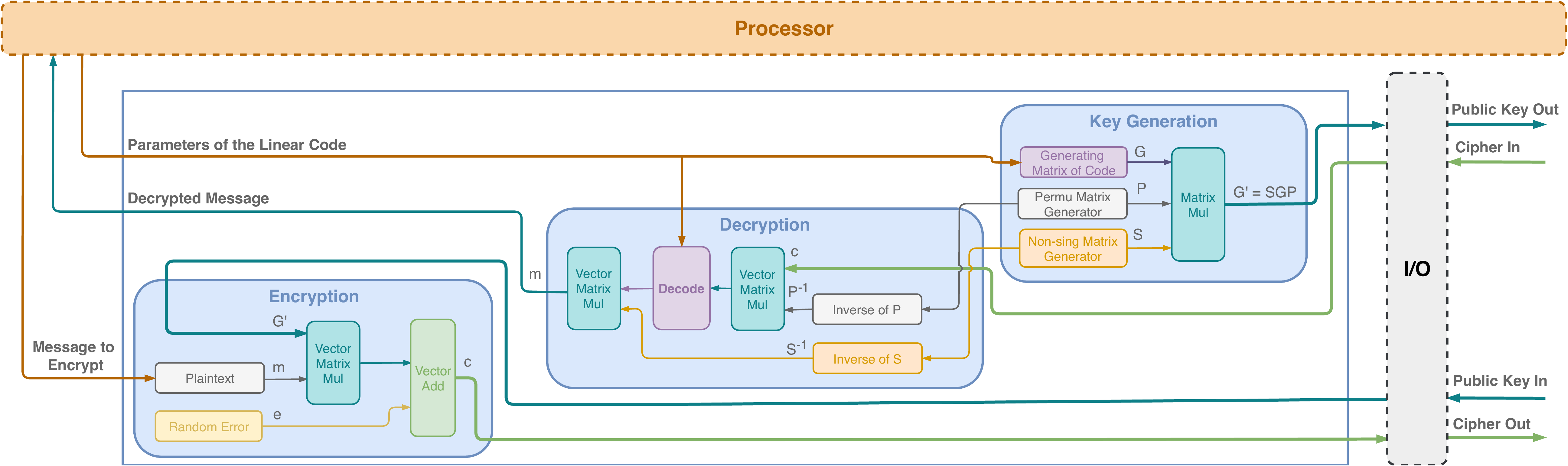} 
	\end{center}
	\vspace{-0.05in}
	\caption{McEliece Cryptosystem Co-Processor Architecture.}
	\label{fig:co-processor}
	\vspace{-0.1in}
\end{figure*}

\subsection{The McEliece Algorithm} \label{sec: PKC}
\vspace{-0.05in}
The detailed protocol of the public-key cryptosystem (PKC) can be found in \cite{ME1978}. Here we provide a brief introduction to aid the presentation of the co-processor. 
	
	\textbf{Key generation:} 
	Alice picks a binary $(n, k, t)$ ECC code $C$ with $k$ information (plaintext) bits, $n$ total codeword length, and the capability of correcting up to $t$ random errors. The $k \times n$ generating matrix of $C$ is denoted by $G$. Alice also picks a $k \times k$ binary non-singular matrix $S$ and a $n \times n$ permutation matrix $P$. Then Alice computes the public key $G'$ as: 
	\vspace{-0.05in}
	\begin{equation} \label{eq: keygen}
		G' = SGP.
	\vspace{-0.05in}
	\end{equation}
	Alice keeps $S, G, P$ as the private key. 
	
	\textbf{Encryption:} Bob converts his message (plaintext) into a $k$-bit binary vector $m$, and generates the $n$-bit cipher $\{c\}$ as:
	\vspace{-0.05in}
	\begin{equation} \label{eq: enc}
		c = mG' + e,
	\vspace{-0.05in}
	\end{equation}
	where $e$ is a binary vector weight $t$. 
	
	\textbf{Decryption:} Alice decrypts the cipher by performing:
	\begin{equation} \label{eq: dec}
		m = (\text{Decode}(cP^{-1}))S^{-1}
	\vspace{-0.05in}
	\end{equation}
	where Decode() stands for the error correction function of $C$, and $P^{-1}, S^{-1}$ are the inverse matrices of $P, S$ respectively. 

\vspace{-0.05in}
\subsection{Cryptosystem Co-Processor Architecture} \label{sec:Arch}
The McEliece based public key encryption cryptosystem co-processor has three modules: the \textit{Key Generation}, \textit{encryption} and \textit{decryption}, shown in Fig~\ref{fig:co-processor}. Of the three modules, 
the decryption unit has the highest complexity and consumes the most hardware resources, especially, its \textit{Decode} stage. Therefore, our efficient implementation primarily targets the \textit{Decode} sub-module. 

\noindent 
\textbf{The Orthogonal Latin Square Code (OLSC):}
The OLSC technique is a $t$-error-correcting binary code with its decoding matrix $H = [M | I]$, where $M$ consists of $2t - 2$ orthogonal Latin squares sized $q \times q$, and $I$ is an identity matrix of order $2tq$. Its generating matrix is $G = [I | M^\top]$, where $^\top$ stands for transpose. Once constructed, the columns of generating matrices can be permuted to create different OLSCs with the same parameters. 

Algorithm \ref{alg:OLSC} is the decoding procedure of OLSC codes, which consists mostly of binary linear operations. Thus, it can be carried out fast and efficient in hardware. In \cite{LB2016}, authors generated the binary OLSC to non-binary codes while still maintaining low decoding complexity. By replacing the binary Goppa code with the non-binary OLSC, the decoding stage only requires (1) binary vector-matrix multiplication and (2) $k$ parallel majority votings among $q$ non-binary vectors each. This feature enables much faster decryption time than the Goppa-code based scheme, as shown in Table~\ref{tab:analysis}.

Due to this simpler decoding mechanism, in the OLSC-based McEliece cryptosystem co-processor, we are 
able to design a single-cycle (one-step) decoding unit. This design is much faster than an equivalent binary 
Goppa code-based system. The non-binary OLSC is able to encode a $kb$-bit plaintext with a $k \times n$ generating matrix $G$ ($b$ being the size of each non-binary symbol), while binary Goppa code can only deal with $k$-bit plaintexts with such a matrix. In other words, given the same size of plaintext, the key size of the non-binary OLSC-based McEliece cryptosystem is $1/b$ that of the Goppa code-based.  Given a proper verification on the security level of the OLSC-based scheme with various decoding techniques, it could serve as an efficient and high speed variant of the McEliece cryptosystem. 

\vspace{-0.1in}
\begin{lstlisting} [caption={{OLSC-based McEliece Cryptosystem}}, escapeinside={(*}{*)}, label=alg:OLSC]
(*  Let $G' = SGP$ and $t$ be the public key, and $\{G, S, P\}$ the private key, where $G$ is a $k \times n$ OLSC encoding matrix with random permutation of columns, and $H$ as its corresponding decoding matrix. Let each Latin square be of size $q \times q$, $m$ be the plaintext, and $c$ the encrypted cipher.  *)

Precompute: (*$S^{-1}, P^{-1}$ as the inverse to $S, P$.*)

(*$ c' \gets cP^{-1} $*)
(*$ u \gets Hc' \times H $*)
for i=0 to n
	(*$ m'_i \gets (u_i > q/2) ? \sim c'_i : c'_i  $*)
(*$ m \gets m'S^{-1} $*)
return (*$m$*)
\end{lstlisting} 

\renewcommand{\arraystretch}{1.5}
\begin{table}[htp]  
\vspace{-0.15in}
	\centering
	\small
	\begin{threeparttable}
		\caption{{Complexity Analysis} \label{tab:estimation} \vspace{-0.15in}}
		\begin{tabular}{ c | c  c }
			\hline 
												& Finite Field Ops 			& Latency (cycles)	  	\\
			\hline 
			\hline 
			\textbf{Binary Goppa Code-based} 	& $O(n^2)$  				& $O(n)$  		 		\\
			\hline 
			\textbf{OLSC-based}  				& 0							& $O(1)$			     \\
			\hline
		\end{tabular}
		\label{tab:analysis}
	\end{threeparttable}
	\vspace{-0.15in}
\end{table}

\vspace{-0.1in}
\bibliographystyle{ACM-Reference-Format}
\bibliography{paper} 

\end{document}